\documentclass[11pt]{article}

\usepackage{indentfirst}
\usepackage{graphicx}
\usepackage{longtable}
\usepackage{supertabular}
\usepackage{amsmath}
\usepackage{amssymb}
\usepackage{color}
\usepackage{amstext}
\usepackage{cases}
\usepackage{float}
\usepackage{threeparttable}
\usepackage{booktabs}
\usepackage[margin=.8in]{geometry}
\usepackage{setspace}
\usepackage{multirow}
\usepackage{array}

\begin{document}

\title{{\Large\bf {Measuring knowledge for recognition and knowledge entropy}} }

\author{Fujun Hou \thanks{Tel.: +86 10
6891 8960; Fax: +86 10 6891 2483; email: houfj@bit.edu.cn.} \\
School of Management and Economics\\
Beijing Institute of Technology\\
Beijing, China, 100081\\
}
\date{\today}
\maketitle

\begin{abstract}
People employ their knowledge to recognize things. This paper is concerned with how to measure people's knowledge for recognition and how it changes. The discussion is based on three assumptions. Firstly, we construct two evolution process equations, of which one is for uncertainty and knowledge, and the other for uncertainty and ignorance. Secondly, by solving the equations, formulas for measuring the levels of knowledge and the levels of ignorance are obtained in two particular cases. Thirdly, a new concept \emph{knowledge entropy} is introduced. Its similarity with Boltzmann's entropy and its difference with Shannon's Entropy are examined. Finally, it is pointed out that the obtained formulas of knowledge and knowledge entropy reflect two fundamental principles: (1) The knowledge level of a group is not necessarily a simple sum of the individuals' knowledge levels; and (2) An individual's knowledge entropy never increases if the individual's thirst for knowledge never decreases.

\end{abstract}

\setlength{\unitlength}{1mm}

\section{Introduction}

It is widely agreed that people employ knowledge to recognize things. The term of 'knowledge' can be defined in many ways. In this paper, knowledge refers to the understanding of someone or something, and it is related to the capacity of acknowledgement in human beings \cite{Cavell2002}. When people have full knowledge about the objects of their concerned, they will differentiate objects well from each other. If one has no knowledge at all, he/she will not make any distinction on the objects. In other cases (only with part of knowledge), one may show some uncertainty which means that he/she can not distinguish some objects but think that they are similar. Although knowledge is virtually an abstract concept, but people do have always depended on their knowledge to make distinctions among the objects under consideration. A question thus arises: How to mathematically measure people's knowledge for recognition and how does it change? This paper aims to answer this question.

Our discussion is based on some assumptions. The rationality of the assumptions is illustrated by some famous quotes. This is presented in Section 2. We investigate the influence of knowledge on uncertainty and the influence of ignorance on uncertainty in Section 3. The model includes two first-order ordinary differential equations and their general solution formulas. Particular boundary conditions for two particular cases and their corresponding formulas for measuring knowledge and ignorance are discussed and presented in Section 4. In Section 5, we introduce the expression of knowledge entropy by comparing the obtained ignorance formula and Boltzmann's entropy formula. The properties of knowledge entropy are examined and two principles are thus deduced in Section 6. In Section 7, we point out a main difference between the knowledge entropy and the Shannon Entropy. Section 8 concludes the paper.

\section{Notations and assumptions}

Knowledge makes a great help for recognition as a motto puts it "Knowledge enables you to add a pair of eyes". There are also two other terms related to knowledge, of which one is the ignorance, and the other is the uncertainty. Ignorance means a lack of knowledge. Good knowledge results in good recognition. On the contrary, absence of knowledge means ignorance and often leads to uncertainty in recognition. In our opinion, knowledge, ignorance and uncertainty are all variables that can be measured mathematically.

We will use the following notations:
\begin{itemize}

\item[$\diamond$] $K$: the level of knowledge for recognition.

\item[$\diamond$] $I$: the level of ignorance.

\item[$\diamond$] $U$: the level of uncertainty as a result of the absence of knowledge.

\item[$\diamond$] $|A|$: cardinal number of set $A$.

\item[$\diamond$] $X$: a set of objects, $X=\{x_1,x_2,\ldots,x_n\}$, $1<n<+\infty$.

\item[$\diamond$] $\sim$: an equivalence relation on $X$.

\item[$\diamond$] $\preceq$: a weak order relation on $X$.

\end{itemize}
When we assume that $K$, $I$ and $U$ are measurable variables, then these three variables are bounded ones. Without loss of generality, we assume $0\leq K\leq 1$ and $0\leq I\leq 1$. We also write $U_{\min}\leq U\leq U_{\max}$.

Moreover, to conduct our discussion, the following assumptions are needed:
\begin{itemize}

\item[]\textbf{Assumption I:} People are in enthusiasm for acquiring knowledge to get to know the world around them, and the process of acquiring knowledge is a continuous process.

\item[]\textbf{Assumption II:}  The more knowledge the people possess, the less uncertainty will the people's recognitions have.

\item[]\textbf{Assumption III:}  The more the uncertainty of the people's recognition has, the more likely the people will seek knowledge so as to make a better recognition.

\end{itemize}

The above assumptions are rational for ordinary people because they are compatible with some famous quotes:
\begin{itemize}

\item[(1)] "All men by nature desire knowledge" (Aristotle (384 BC - 322 BC), Metaphysics), and "The desire of knowledge, like the thirst of riches, increases ever with the acquisition of it" (Laurence Sterne, 1713-1768).

\item[(2)] "Knowledge is power" (Sir Francis Bacon (1561 - 1626)).

\item[(3)] "To be conscious that you are ignorant is a great step to knowledge" (Benjamin Disraeli (1804 - 1881)), "Perplexity is the beginning of knowledge" (Kahlil Gibran (1883 - 1931)), and "The beginning of knowledge is the discovery of something we do not understand" (Frank Herbert (1920 - 1986)).

\end{itemize}

\section{Models}

According to Assumptions I and II, the influence of knowledge on uncertainty can be described by a formula
$$\frac{dU}{dK}<0.$$
Similarly, the influence of ignorance on uncertainty can be described by
$$\frac{dU}{dI}>0.$$

According to Assumption III, the relation of knowledge and uncertainty, as well as the relation of ignorance and uncertainty can be (but not have to be) further written as
$$\frac{dU}{dK}=aU,\eqno(1)$$
and
$$\frac{dU}{dI}=bU,\eqno(2)$$
respectively, where $a$ and $b$ are two non-zero constants.

Formulas (1) and (2) are two first-order ordinary differential equations. Their general solution formulas are given as follows:
$$\ln U=aK+c_1, \eqno(3)$$
and
$$\ln U=bI+c_2,\eqno(4)$$
respectively, where $c_1$ and $c_2$ are two constants.

In particular cases, if the boundary conditions were known, then their solutions could be known as well. We will discuss this in next section. We \emph{remark} that Eqs. (3) and (4) represent continuous processes. In the following sections, however, we will consider some particular cases where $U$ will take integer values.

\section{Measuring knowledge in two particular cases}

Equivalence relation and weak order relation are two basic relations, which are closely related to human cognition. Living in the society, people have to make out things and make optimal choices. When making out things, one may say "This object is equal to that one" or "These two objects are different", and so on. This kind of sayings indicate that people made distinctions under an equivalence relation even if they might not realize this. When making optimal choices, people also use their knowledge to express preferences, such as "These alternatives are indifferent to me" or "This alternative is more preferred to me than that one", and so on. Even though we cannot say someone's preference is good while the other's is bad, however, people do use their knowledge to discern objects in their preferences. Frequently, people use the equivalence relations to differentiate things, and the weak order relations to express preferences. In this section, we consider how to measure the knowledge for recognition (particularly, for classification) with respect to an equivalence relation and a weak order relation that are both constructed on finite-element sets.

\subsection{When making distinctions under equivalence relations}

An equivalence relation over an object set corresponds to a partition of the object set, and vice versa. The partition is composed of disjoint equivalence classes. Two objects are equivalent to each other if and only if they belong to the same equivalence class \cite{Wilder1965}.

Consider an equivalence relation $\sim$ which is defined on object set $X=\{x_1,x_2,\ldots,x_n\}$. $x_i\sim x_j$ means that $x_i$ and $x_j$ are equivalent to each other. Let $[x_i]$ be the equivalence class corresponding to $x_i$, where $[x_i]=\{x_j\mid x_j\in X, x_j\sim x_i\}$.  Clearly, if we have full knowledge about objects then we will differentiate any one from others. In this case, one equivalence class will include only one object; on the contrary, if we have no knowledge at all then we will differentiate none from each others. In this case, all the objects will be included in one equivalence class. Therefore, the cardinal numbers of the equivalence classes reflect a relationship of the knowledge for recognition and the uncertainty. Mathematically, if we define
$$U=W=\sum\limits_{i=1}^{n}\mid[x_i]\mid,\eqno(5)$$
then we have $W_{\min}=n$ and $W_{\max}=n^2$, and the minimum value and the maximum value correspond to the 'full knowledge (hence no uncertainty)' case and the 'no knowledge (hence full uncertainty)' case, respectively. In this case the boundary condition of formula (3) can be
$$
\begin{cases}
K=1, & \text{if $U=W=n$},\\
K=0, & \text{if $U=W=n^2$}.
\end{cases}\eqno(6)
$$
Thus formula (3) is changed into
$$(-\ln n) K+\ln n^2=\ln W,$$
namely,
$$K=\frac{1}{\ln n}\ln \frac{n^2}{W}.\eqno(7)$$

Similarly, the boundary condition of formula (4) can be
$$
\begin{cases}
I=0, & \text{if $U=W=n$},\\
I=1, & \text{if $U=W=n^2$}.
\end{cases}\eqno(8)
$$
Under this condition, formula (4) is changed into
$$(\ln n)I+\ln n=\ln W,$$
namely,
$$I=\frac{1}{\ln n}\ln \frac{W}{n}.\eqno(9)$$

By synthesizing Eqs.(7) and (9) we have
$$K+I=1.\eqno(10)$$

Because $n\leq W\leq n^2$, thus we have $K\in [0,1]$, $I\in [0,1]$. Formula (10) implies a complementary relationship between $K$ and $I$, which indicates "the more knowledge, the less ignorance".

\vspace{0.2cm}
For illustrative purpose, we consider an example.

\textbf{Example 1} Suppose that there were 5 eggs, $X=\{egg1,egg2,egg3,egg4,egg5\}$, and that two persons, $\{John, Jack\}$, were asked to use their individual scales to check the weights of the eggs. They reported their weighing results as shown by Table 1.

\begin{table}[h]
\small{\renewcommand{\thetable}{1} \caption{\label{tab:test}Weighing results} \vskip
2mm
\begin{center}
\begin{threeparttable}
\begin{tabular}{llll}
 \hline
 & egg$^a$ & John &
Jack \\
\hline
& egg1 & 60 & 60.2\\
& egg2 & 63 & 62.8\\
& egg3 & 63 & 63.1\\
& egg4 & 61 & 61.1\\
& egg5 & 61 & 61.1\\
 \hline
\end{tabular}
\footnotesize{$^a$ Weighed in grams.}\\
\end{threeparttable}
\end{center}}
\end{table}

We use formula (7) to measure the two persons' knowledge levels in their recognitions of the eggs' weights.

The equivalence classes deduced from John's weighing results are:
\begin{equation*}
\begin{aligned}
&[egg1]=\{egg1\},\\
&[egg2]=\{egg2,egg3\}, [egg3]=\{egg2,egg3\}, \\
&[egg4]=\{egg4,egg5\}, \mbox{and~} [egg5]=\{egg4,egg5\}.
\end{aligned}
\end{equation*}

By using formula (7), John's knowledge level is measured as
    $$K^{(John)}=\frac{1}{\ln 5}\ln \frac{5^2}{1+2+2+2+2}=\frac{1}{\ln 5}\ln \frac{25}{9}\doteq 0.6348.$$

The equivalence classes deduced from Jack's weighing results are:
\begin{equation*}
\begin{aligned}
&[egg1]=\{egg1\},\\
&[egg2]=\{egg2\}, \\
&[egg3]=\{egg3\}, \\
&[egg4]=\{egg4,egg5\}, \mbox{and~} [egg5]=\{egg4,egg5\}.
\end{aligned}
\end{equation*}

By using formula (7), Jack's knowledge level is measured as
    $$K^{(Jack)}=\frac{1}{\ln 5}\ln \frac{5^2}{1+1+1+2+2}=\frac{1}{\ln 5}\ln \frac{25}{7}\doteq 0.7909.$$

One can see that Jack has a higher knowledge level than John. This result is consistent with our intuitive perception of the data shown by Table 1.

\subsection{When expressing preferences under weak order relations}

Consider a weak order relation $\preceq$ which is defined on object set $X=\{x_1,x_2,\ldots,x_n\}$. On the one hand, if one thinks that $x_i$ and $x_j$ are indifferent and thus a tie arises, then we write $x_i\sim x_j$. On the other hand, if one thinks that $x_i$ is preferred to $x_j$, then we write $x_i\succ x_j$ \cite{Roberts2011}. We assume that people use ties-permitted ordinal rankings to express their preferences.

A ties-permitted ordinal ranking of a set can be represented by a preference sequence whose entries are sets containing possible ranking positions of objects \cite{Hou2015a,Hou2015b}. A preference sequence $PS=(PS_i)_{n\times 1}$ is defined by
$$PS_i=\{|\xi_i|+1,|\xi_i|+2,\ldots,|\xi_i|+|\eta_i|\},\eqno(11)$$
where $\xi_i=\{x_k\mid x_k\in X,x_k\succ x_i\}$ and $\eta_i=\{x_k\mid x_k\in X,x_i\sim x_k\}$.

For instance, let $X=\{x_1,x_2,x_3,x_4\}$ and a weak ordering on the set be: $x_1\succ x_2\sim x_3\succ x_4$, then the corresponding preference sequence is as follows:
$$PS=(PS_i)_{4\times 1}=(\{1\},\{2,3\},\{2,3\},\{4\})^T.$$

Evidently, different entries of a preference sequence constitute a partition of the set $\{1,2,\ldots,n\}$ \cite{Hou2015a}. This is understandable because the elements in an entry are deduced based on the relation $x_i\sim x_k$ as shown by formula (11), which is in nature an equivalence relation. Therefore, the results obtained in subsection 4.1 can be directly applied to the discussion in this subsection.

We define the cardinal number of a preference sequence, $PS=(PS_i)_{n\times 1}$, as follows:
$$\mid PS\mid=\sum\limits_{i=1}^{n}\mid PS_i\mid.\eqno(12)$$
Clearly, we have $\mid PS\mid_{\min}=n$ and $\mid PS\mid_{\max}=n^2$, and the larger the cardinal number of a person's preference sequence is, the more distinct the objects will be to the person. Therefore, the knowledge for recognition in a person's preference sequence can also be measured by formula (7), where we need only to substitute $W$ by $\mid PS\mid$. That is
$$K_{PS}=\frac{1}{\ln n}\ln \frac{n^2}{\mid PS\mid}.\eqno(13)$$

Similar to formula (9), the ignorance in a person's preference sequence can be measured by
$$I_{PS}=\frac{1}{\ln n}\ln \frac{\mid PS\mid}{n}.\eqno(14)$$

Because $n\leq\mid PS\mid\leq n^2$ we also have $K_{PS}\in [0,1]$, $I_{PS}\in [0,1]$ and $K_{PS}+I_{PS}=1$.

\vspace{0.2cm}
Next we consider an example.

\textbf{Example 2} Suppose that there were an alternative set $\{x_1,x_2,x_3,x_4,x_5\}$, and a group of experts, $\{Expert1,Expert2,Expert3\}$. Assume that the experts provided their weak orderings on the alternative set as follows.
\begin{itemize}

\item[] Expert1: $x_1\succ x_2\sim x_3\succ x_4\sim x_5$.

\item[] Expert2: $x_1\succ x_2\sim x_3\sim x_4\succ x_5$.

\item[] Expert3: $x_1\sim x_2\succ x_3\succ x_4\succ x_5$.

\end{itemize}

We use formula (13) to measure the experts' knowledge levels for recognition in their preferences. We first write out their preference sequences according to formula (11):
$$
\begin{array}{c}{\hspace{1cm}\begin{array}{cccccc}Expert1
&\hspace{0.4cm}Expert2 &\hspace{0.4cm}Expert3
\end{array}}\\
{\begin{array}{cc}
{\begin{array}{c} A_1\\
A_2\\
A_3\\
A_4\\
A_5
\end{array}}&
{ \left[\begin{array}{c}
\{1\}\\
\{2,3\}\\
\{2,3\}\\
\{4,5\}\\
\{4,5\}\end{array}\right],
\left[\begin{array}{c}\{1\}\\
\{2,3,4\}\\
\{2,3,4\}\\
\{2,3,4\}\\
\{5\}\end{array}\right], \left[\begin{array}{c}\{1,2\}\\
\{1,2\}\\
\{3\}\\
\{4\}\\
\{5\}\end{array}\right]
}
\end{array}.}
\end{array}
$$

By using formula (13), the experts' knowledge levels are measured as
\begin{itemize}

\item[] Expert1: $K_{PS}^{(E1)}=\frac{1}{\ln 5}\ln \frac{5^2}{1+2+2+2+2}\doteq 0.6348$.

\item[] Expert2: $K_{PS}^{(E2)}=\frac{1}{\ln 5}\ln \frac{5^2}{1+3+3+3+1}\doteq 0.5101$.

\item[] Expert3: $K_{PS}^{(E3)}=\frac{1}{\ln 5}\ln \frac{5^2}{2+2+1+1+1}\doteq 0.7909$.

\end{itemize}

\section{Knowledge entropy}

In thermodynamics and statistical mechanics, there is a well-known formula called Boltzmann's entropy formula
$$S=k_B\ln \Omega.\eqno(15)$$
In the above formula, $S$ is the entropy, $k_B$ is Boltzmann constant, and $\Omega$ is the number of microstates consistent with the given equilibrium macrostate.

If we rewrite the formula of ignorance, namely, formula (9), as
$$I=\frac{1}{\ln n}\ln \frac{W}{n}=\frac{1}{\ln n}\ln {W} -1$$
hence
$$I+1=\frac{1}{\ln n}\ln {W},$$
then we have an expression that is quite similar to Boltzmann's entropy formula. Thus we introduce a concept \emph{knowlege entropy}, denoted $S_K$, as follows
$$S_K=\frac{1}{\ln n}\ln {W}.\eqno(16)$$
and we have
$$S_K=I+1.\eqno(17)$$

Similarly, we can also define another kind of knowlege entropy based on formala (14), as in the following
$$S_{Kps}=\frac{1}{\ln n}\ln {|PS|}.\eqno(18)$$
and we also have
$$S_{Kps}=I_{PS}+1.\eqno(19)$$

From formulas (10) and (17),  we have
$$K=K-0=1-I=2-S_k.$$
Thus we obtain $K-0=2-S_k$, which can be further written as
$$K-\frac{1}{\ln n}\ln \frac{n^2}{n^2}=\frac{1}{\ln n}\ln {n^2}-S_k.$$
Taking into account that $W_{\max}=n^2$ and plugging formulas (7) and (16) into the above expression, we have
$$K(W)-K(W_{\max})=S_K(W_{\max})-S_K(W).\eqno(20)$$

Formula (20) indicates that "acquiring knowledge means the decrease of knowledge entropy".

Further, from formula (18) we know that "the lower is the knowledge entropy, the more distinctive is the people's ranking order", since a lower entropy indicates a lower cardinal number of a preference sequence hence a more distinctive ranking.

We illustrate the usage of the knowledge entropy formula when the data of Example 2 are used. From $K_{PS}+I_{PS}=1$ and $S_{Kps}=I_{PS}+1$, we have $S_{Kps}=2-K_{PS}$. We have obtained the knowledge levels in Example 2. Therefore, we can calculate the knowledge entropies of the experts' preferences:
\begin{itemize}

\item[] Expert1: $S_{Kps}^{(E1)}=2-K_{PS}^{(E1)}\doteq 1.3652$.

\item[] Expert2: $S_{Kps}^{(E2)}=2-K_{PS}^{(E2)}\doteq 1.4899$.

\item[] Expert3: $S_{Kps}^{(E3)}=2-K_{PS}^{(E3)}\doteq 1.2091$.

\end{itemize}
The above results indicate that expert 3 had the lowest level of entropy and has made the most distinct recognition of the alternatives.

\section{Two deduced principles}

Taking into account Assumption I "People are in enthusiasm for acquiring knowledge to get to know the world around them" and the implication of formula (20) "Acquiring knowledge means the decrease of knowledge entropy", we have the following deduction:

\vspace{0.2cm}
\textbf{Principle 1} An individual's knowledge entropy never increases if the individual's thirst for knowledge never decreases.

\vspace{0.2cm}

Next we will introduce another principle by examining the additivity of the knowledge measure.

\textbf{Proposition 1} The knowledge measure expressed by formula (7) (or formula (13)) does not fulfil the property of sub-additivity.

\textbf{Proof} The above proposition has two implications:
\begin{itemize}

\item The overall knowledge of two persons is not necessarily the sum of their individual knowledge.

\item For one person, his/her knowledge on a set of objects is not necessarily the sum of his/her knowledge on the sub-sets.

\end{itemize}
We use counter examples to illustrate.
\begin{itemize}

\item[-] Suppose that the object set is $\{x_1,x_2,x_3\}$. Thus we have $n=3$ and in this case $3\leq W\leq 3^2$.

Suppose that two persons Alan and Barbara were asked to discern the objects.

Assume that Alan said that "$x_1$ is different from others, while $x_2$ and $x_3$ are equivalent". Hence the equivalence classes deduced from Alan's knowledge for recognition are
    $$[x_1]=\{x_1\}, [x_2]=\{x_2,x_3\}, \mbox{and~} [x_3]=\{x_2,x_3\}.$$
    By using formula (7), Alan's knowledge is measured as
    $$K^{(A)}=\frac{1}{\ln 3}\ln \frac{3^2}{1+2+2}=\frac{1}{\ln 3}\ln \frac{9}{5}.$$

    Assume that Barbara said that "the objects are all different from each other". Hence the equivalence classes deduced from Barbara's knowledge for recognition are
    $$[x_1]=\{x_1\}, [x_2]=\{x_2\}, \mbox{and~} [x_3]=\{x_3\}.$$
    By using formula (7), Barbara's knowledge is measured as
    $$K^{(B)}=\frac{1}{\ln 3}\ln \frac{3^2}{1+1+1}=\frac{1}{\ln 3}\ln \frac{9}{3}.$$

    Let $K^{(A+B)}=K^{(A)}+K^{(B)}$, namely, $\frac{1}{\ln 3}\ln \frac{9}{W}=\frac{1}{\ln 3}\ln \frac{9}{5}+\frac{1}{\ln 3}\ln \frac{9}{3}$.

    However, we cannot find a $W$ satisfying both $3\leq W\leq 3^2$ and $\frac{1}{\ln 3}\ln \frac{9}{W}=\frac{1}{\ln 3}\ln \frac{9}{5}+\frac{1}{\ln 3}\ln \frac{9}{3}$. Indeed, the only solution of the equation $\frac{1}{\ln 3}\ln \frac{9}{W}=\frac{1}{\ln 3}\ln \frac{9}{5}+\frac{1}{\ln 3}\ln \frac{9}{3}$ is $W=\frac{15}{9}$, but beyond $3\leq W\leq 3^2$. Therefore, the first implication is verified.

\item[-] Suppose that the object set is $\{x_1,x_2,x_3,x_4\}$. Thus we have $n=4$ and in this case $4\leq W\leq 4^2$.

Suppose that a person Cassie was asked to discern the objects. Assume that Cassie said that "$x_1$ is different from others, so does $x_4$, while $x_2$ and $x_3$ are equivalent". Hence the equivalence classes deduced from Cassie's knowledge for recognition are
    $$[x_1]=\{x_1\}, [x_2]=\{x_2,x_3\}, [x_3]=\{x_2,x_3\}, \mbox{and~} [x_4]=\{x_4\}.$$
    By using formula (7), Cassie's knowledge is measured as
    $$K^{(C)}=\frac{1}{\ln 4}\ln \frac{4^2}{1+2+2+1}=\frac{1}{\ln 4}\ln \frac{16}{6}.$$

    Now we divide $X$ into two sub-sets: $X_1=\{x_1,x_4\}$ and $X_2=\{x_2,x_3\}$. By using formula (7), Cassie's knowledge on these two sub-sets is measured as
    $$K^{(X_1)}=\frac{1}{\ln 2}\ln \frac{2^2}{1+1}=\frac{1}{\ln 2}\ln \frac{4}{2}=1,$$
    and
    $$K^{(X_2)}=\frac{1}{\ln 2}\ln \frac{2^2}{2+2}=\frac{1}{\ln 2}\ln \frac{4}{4}=0,$$
    respectively. Evidently, we have $K^{(C)}\neq K^{(X_1)}+K^{(X_2)}$. Therefore, the second implication is verified and hence the proof is finished.$\Box$

\end{itemize}

\vspace{0.2cm}
From Proposition 1 we can deduce another principle as follows.

\vspace{0.2cm}
\textbf{Principle 2} The knowledge level of a group  is not necessarily a simple sum of the individuals' knowledge levels.

\section{A main difference of knowledge entropy and Shannon entropy}

Because the knowledge measure (formula (7)) and the knowledge entropy measure (formula (16)) have a relationship of $K=2-S_K$, namely $K+S_K=2$, we thus have a result from Proposition 1 as follows.

\textbf{Proposition 2} The knowledge entropy does not fulfil the property of sub-additivity.

\vspace{0.2cm}
Shannon entropy, also called information entropy, was introduced by Claude Shannon \cite{Shannon1948}. It is expressed as
$$S=-\sum\limits_{i=1}P_i\ln P_i.$$
It is known that Shannon entropy satisfies an additivity property, namely, if two subsets are disjoint, then the total entropy of the two subsets will be the sum of their individual entropies of the two subsets.

However, as shown by Proposition 2, the knowledge entropy introduced in this paper does not have the sub-additivity property. This is a main difference of the knowledge entropy and Shannon entropy.

\section{Conclusion}

We have discussed how to measure the knowledge for recognition (particularly, for classification) and how it changes. Our discussion were based upon three assumptions. We proposed an equation for describing the influence of knowledge on uncertainty, and obtained two particular formulas for measuring the levels of knowledge for recognitions in two particular cases. Moreover, by investigating the evolution process of ignorance with uncertainty, a concept \emph{knowledge entropy} was introduced and its formula was presented. Its similarity with Boltzmann's entropy and its difference with Shannon's entropy were examined. Furthermore, based on a mathematical analysis, we obtained or evidenced the following results:
\begin{itemize}

\item[(1)] Acquiring knowledge means the decrease of knowledge entropy.

\item[(2)] The lower is the knowledge entropy, the more distinctive is the people's ranking order.

\item[(3)] The knowledge level of a group is not necessarily a simple sum of the individuals' knowledge levels.

\item[(4)] An individual's knowledge entropy never increases if the individual's thirst for knowledge never decreases.

\end{itemize}

\vspace{0.3cm}
 \textbf{Acknowledgments}

The work was supported by the National Natural Science Foundation
of China (No. 71571019).

\end{document}